\input harvmac.tex
\let\includefigures=\iftrue
\newfam\black
\includefigures
\input epsf
\def\figin{\epsfcheck\figin}\def\figins{\epsfcheck\figins}
\def\epsfcheck{\ifx\epsfbox\UnDeFiNeD
\message{(NO epsf.tex, FIGURES WILL BE IGNORED)}
\gdef\figin##1{\vskip2in}\gdef\figins##1{\hskip.5in}
\else\message{(FIGURES WILL BE INCLUDED)}%
\gdef\figin##1{##1}\gdef\figins##1{##1}\fi}
\def\DefWarn#1{}
\def\figinsert{\goodbreak\midinsert}
\def\ifig#1#2#3{\DefWarn#1\xdef#1{fig.~\the\figno}
\writedef{#1\leftbracket fig.\noexpand~\the\figno}%
\figinsert\figin{\centerline{#3}}\medskip\centerline{\vbox{\baselineskip12pt
\advance\hsize by -1truein\noindent\footnotefont{\bf Fig.~\the\figno:} #2}}
\bigskip\endinsert\global\advance\figno by1}
\else
\def\ifig#1#2#3{\xdef#1{fig.~\the\figno}
\writedef{#1\leftbracket fig.\noexpand~\the\figno}%
\global\advance\figno by1}
\fi

\Title{\vbox{\baselineskip12pt\hbox{hep-th/0006160}
\hbox{CALT-68-2283}\hbox{CITUSC/00-027}}}
{\vbox{
\centerline{Quantum Field Theories With Compact}
\vskip 8pt
\centerline{Noncommutative Extra Dimensions}
}}
\centerline{Jaume Gomis, Thomas Mehen and Mark B. Wise}
\centerline{\it Department of Physics, California Institute of Technology, Pasadena, CA 91125}
\centerline{\it and}
\centerline{\it Caltech-USC Center for Theoretical Physics} 
\centerline{\it University of Southern California}
\centerline{\it Los Angeles, CA 90089}

\centerline{\tt gomis,\ mehen,\ wise@theory.caltech.edu}
\medskip
\noindent



We study field theories on spaces with additional compact noncommutative
dimensions. As an example, we study $\phi^3$ on
$R^{1,3}\times T^{2}_\theta$ using perturbation theory.
The infrared divergences in the noncompact theory give rise to unusual
dynamics for the mode of $\phi$ which is constant along the torus. 
Correlation functions involving this mode vanish. Moreover, 
we show that the spectrum of Kaluza-Klein excitations can be very different from the 
analogous commuting theory.  There is an additional contribution to the Kaluza-Klein 
mass formula that resembles the contribution of winding states in string theory. We also 
consider the effect of noncommutativity on the four dimensional Kaluza-Klein excitations 
of a six dimensional gauge field.


\Date{June 2000}

\newsec{Introduction}

Quantum field theories on noncommutative spaces -- usually referred to
as noncommutative field theories -- generalize the familiar structure of
conventional (commutative) local quantum field theories. Field
theories on these spaces can be studied using conventional  
techniques by representing the underlying noncommutative
structure that defines the noncommutative space in terms of more
familiar entities in a commutative space. For example, noncommutative
Minkowski space $R_\theta^{1,d}$ is defined in terms of space-time
coordinates $x^\mu, \mu=0,\ldots,d$, which satisfy the following
commutation relations
\eqn\comm{
[x^\mu,x^\nu]=i \,\theta^{\mu\nu}\qquad \mu,\nu=0,\ldots,d,}
where $\theta^{\mu\nu}$ is a real antisymmetric
matrix. 
$R_\theta^{1,d}$ is a noncommutative associative algebra, with
elements given by ordinary continuous functions on $R^{1,d}$ and whose
product is given by the Moyal bracket or $\star$-product
of functions
\eqn\produ{
(f \star g) (x)= e^{{i
\over 2} \theta^{\mu\nu} {\partial \over \partial \alpha^\mu}{\partial \over
\partial \beta^\nu}} f(x+\alpha)\,
g(x+\beta)|_{\alpha=\beta=0}=fg+{i\over 2}\theta^{\mu\nu}\partial_\mu
f\partial_\nu g+ {\cal O}(\theta^2).}
The generalization of conventional field theories on
$R^{1,d}$ to $R_\theta^{1,d}$ is achieved by replacing the
usual multiplication of fields in the action by the $\star$-product of fields. 

\nref\mvs{S. Minwalla, M.V. Raamsdonk and N. Seiberg, ``Noncommutative
Perturbative Dynamics'', hep-th/9912072.}%
\nref\vs{M.V. Raamsdonk and N. Seiberg, `` Comments on Noncommutative
Perturbative Dynamics'', hep-th/0002186.}%
\nref\haya{M. Hayakawa,``Perturbative analysis on infrared
and ultraviolet aspects of
noncommutative QED on $R^4$,'' hep-th/9912167.}%
\nref\texa{W. Fischler, E. Gorbatov, A. Kashani-Poor, S. Paban,
P. Pouliot and J. Gomis, ``Evidence for winding states in
noncommutative quantum field theory,'' hep-th/0002067.}%
\nref\sunew{A. Matusis, L. Susskind and N. Toumbas,
``The IR/UV connection in the non-commutative gauge theories,''
hep-th/0002075.}%

The presence of $\star$-products in the action leads to theories that are not
maximally Lorentz invariant nor local. The nonlocality of the theory is apparent from
the infinite number of derivatives that appear in the action. The noncommutativity of
the space-time coordinates \comm\ gives rise to a space-time uncertainty relation 
\eqn\uncert{
\Delta x^\mu\Delta x^\nu\ge {1\over 2}|\theta^{\mu\nu}|.}
A dramatic consequence of this relation is the absence of decoupling
of scales in these theories \mvs-\sunew . For instance, if
$\theta^{12}\neq 0$, short 
distance scales in the $x^1$ direction correspond to  very large
distance scales   in
the $x^2$ direction. Therefore, these theories exhibit a very peculiar
mixing between the ultraviolet and the infrared, such that very high
energy modes have  drastic effects on  low energy processes.  In spite 
of this, the theories are apparently renormalizable
\nref\filk{T. Filk, ``Divergences in a Field Theory on Quantum Space'',
Phys. Lett. {\bf B376} 53 (1996).}%
\nref\vgra{J.C. Varilly and J.M. Gracia-Bondia, ``On the ultraviolet
behaviour of quantum fields over noncommutative manifolds'',
Int. J. Mod. Phys. {\bf A14} (1999) 1305, hep-th/9804001.}%
\nref\two{M. Chaichian, A. Demichev and P. Presnajder, ``Quantum Field
Theory on Noncommutative Space-times and the Persistence of
Ultraviolet Divergences'', hep-th/9812180;  ``Quantum Field Theory on
the Noncommutative Plane with E(q)(2) Symmetry'', hep-th/9904132.}%
\nref\rruiz{C.P. Martin, D. Sanchez-Ruiz,
``The One-loop UV Divergent Structure of U(1) Yang-Mills
 Theory on Noncommutative $R^4$'',
Phys. Rev. Lett. {\bf 83} (1999) 476-479,hep-th/9903077}%
\nref\rjabbari{ M. Sheikh-Jabbari, ``One Loop Renormalizability
of Supersymmetric Yang-Mills Theories on Noncommutative Torus'',
JHEP {\bf 06} (1999) 015, hep-th/9903107; ``Noncommutative Super
Yang-Mills Theories with 8 Supercharges and Brane Configurations'', 
hep-th/0001089.}%
\nref\yaaa{I.Ya. Aref\'eva, D.M. Belov and A.S. Koshelev, ``Two-Loop
Diagrams in Noncommutative $\phi^4_4$ theory'', hep-th/9912075; 
``A Note on UV/IR for Noncommutative Complex Scalar Field'',
hep-th/0001215; ``UV/IR Mixing for Noncommutative Complex Scalar Field
Theory, II (Interaction with Gauge Fields)'', hep-th/0003176.}%
\nref\gkw{H. Grosse, T. Krajewski and R. Wulkenhaar, ``Perturbative
quantum gauge fields on the noncommutative torus'', hep-th/9903187;
``Renormalization of noncommutative Yang-Mills theories: A simple 
example'', hep-th/0001182.}%
\nref\twopfo{S. Cho, R. Hinterding, J. Madore and H. Steinacker,
``Finite Field Theory on Noncommutative Geometries'',
hep-th/9903239.}%
\nref\three{E. Hawkins,``Noncommutative Regularization for the
Practical Man'',hep-th/9908052.}%
\nref\riouri{I. Chepelev and  R. Roiban, 
``Renormalization of Quantum Field Theories on Noncommutative $R^d$, I. 
Scalars,''  hep-th/9911098.}%
\nref\ggrd{H.O. Girotti, M. Gomes, V.O. Rivelles and A.J. da Silva,
``A Consistent Noncommutative Field Theory: the Wess-Zumino Model'',
hep-th/0005272.}%
\nref\gubson{S.S. Gubser and S.L. Sondhi, ``Phase structure of
non-commutative scalar field theories'', hep-th/0006119.}%
\filk-\gubson.
Morever, field theories with only space noncommutativity (that is,
$\theta^{0i}=0$)  
have a unitary S-matrix. On the other hand, theories with only space-time
noncommutativity   (that is $\theta^{0i}\neq 0$) are not unitary
\nref\gm{J. Gomis and T. Mehen, "Space-Time Noncommutative Field Theories
and Unitarity", hep-th/0005129.}%
\gm .

A strong motivation for understanding these field theories is the
appearance they make in string theory
\nref\cds{A. Connes, M.R. Douglas and A. Schwarz, ``Noncommutative
Geometry and Matrix Theory: Compactification on Tori'', JHEP {\bf
9802}(1998) 003, hep-th/9711162.}%
\nref\dh{M.R. Douglas and C. Hull, ``D-branes and the Noncommutative
Torus'', JHEP {\bf 9802} (1998) 008, hep-th/9711165.}%
\nref\seibwitt{N. Seiberg and E. Witten, ``String
Theory and Noncommutative Geometry'', JHEP {\bf 9909} (1999) 032,
hep-th/9908142.}%
\nref\chkr{Y.-K. E. Cheung and M. Krogh, ``Noncommutative Geometry
From 0-Branes In A Background B Field'', Nucl. Phys. {\bf B528} (1998)
185, hep-th/ 98030031.}%
\nref\chuho{C.-S. Chu and P.-M. Ho, ``Noncommutative Open String And
D-brane'', Nucl. Phys. {\bf B550} (1999) 151, hep-th/9812219;
``Constrained Quantization of open string in background B field and
noncommutative D-brane'', hep-th/9906192.}%
\nref\scho{V. Schomerus, ``D-Branes And Deformation Quantization'',
JHEP {\bf 9906:030} (1999), hep-th/9903205.}%
\nref\ararajabb{F. Ardalan, H. Arfaei and M.M. Sheikh-Jabbari, ``Mixed
Branes and M(atrix) Theory on Noncommutative Torus'', hep-th/9803067;
``Noncommutative Geometry From Strings and Branes'', JHEP {bf 9902}
(1999), hep-th/9810072; ``Dirac Quantization of Open Strings and
Noncommutativity in Branes'', hep-th/9906161.}%
\cds-\ararajabb .  For instance, noncommutative
gauge theories with space noncommutativity describe the low energy 
excitations of open strings on D-branes in a background Neveu-Schwarz 
two-form field $B$ \cds\dh\seibwitt . The excited open string states and the 
closed strings decouple and the noncommutative field theory  is the proper 
description of the physics. The lack of covariance of noncommutative field theory 
arises from the expectation value of $B$. The appearance of noncommutative field 
theory in a definite limit of string theory strongly suggests that these field theories 
are sensible quantum field theories. This motivates generalizing the conventional
framework of local quantum field theory in order to understand these theories.

\nref\hierarchy{N. Arkani-Hamed, S. Dimopoulos and G. Dvali, ``The Hierarchy Problem 
and New Dimensions at a Millimeter'', hep-ph/98033115, Phys.\ Lett.\ {\bf B429} (1998) 263;
I. Antoniadis, N. Arkani-Hamed, S. Dimopoulos and G. Dvali, ``New Dimensions at a Millimeter
to a Fermi and Superstrings at a TeV'', Phys.\ Lett.\ {\bf B436} (1998) 257;
I. Antoniadis, ``A Possible New Dimension at a Few TeV'', Phys.\ Lett.\ {\bf B246} (1990) 377;
N. Arkani-Hamed, S. Dimopoulos and J. March-Russell, ``Stabilization of Submillimeter
Dimension: The New Guise of the Hierarchy Problem'', hep-th/9809124;
L. Randall and R. Sundrum, ``A Large Mass Hierarchy from a Small Extra Dimension'',
hep-th/99050221, Phys.\ Rev.\ Lett.\ {\bf 83} (1999) 3370. }

\nref\cc{N. Arkani-Hamed, S. Dimopoulos, N. Kaloper and R. Sundrum, ``A Small
Cosmological Constant from a Large Extra Dimension'', hep-th/00031197, 
Phys.\ Lett.\ {\bf B480} (2000) 193; S. Kachru, M. Schulz and E. Silverstein, 
``Self Tuning Flat Domain Walls in 5-D Gravity and String Theory'', hep-th/0001206.}

In nature we observe three spatial dimensions. However, it is possible that
there are additional spatial dimensions. Such is the case in string theory 
and recently there has been speculation that extra dimensions may play a role 
in understanding the hierarchy puzzle \hierarchy\ or the cosmological constant 
problem \cc . If extra dimensions exist the simplest way to explain why they 
have not been observed is to assume they are compact. It is possible that such 
additional compact spatial dimensions are noncommuting. In this paper we explore 
this possibility in some simple field theories that contain two additional compact 
noncommuting dimensions. We find that the UV-IR mixing that exists for infinite 
noncommuting dimensions leads to some interesting properties of the compactified 
theories.

In this paper we study in perturbation theory scalar and gauge quantum field 
theories with compact noncommutative extra dimensions. We examine the low 
energy dynamics of these theories and their spectrum of Kaluza-Klein excitations. 
For concreteness, we consider six dimensional field theories, such that the two extra 
dimensions are compact and noncommutative and four noncompact dimensions are
commutative. The two extra dimensions are taken to correspond to a noncommutative 
two torus $T^2_\theta$ whose coordinates satisfy
\eqn\tor{
[x^4,x^5]=i \, \theta .}
This system can be realized in string theory by wrapping
a five-brane on a two-torus $T^2$ with a constant $B$-field along the torus. 
The low energy effective four dimensional theory resulting from
compactification on a noncommutative space is local and Lorentz
invariant, hence it can be relevant phenomenologically. 

In section $2$ we consider scalar $\phi^3$ theory on $R^{1,3}\times T_{\theta}^2$. 
The noncommutative theory on a noncompact space contains infrared divergences.
In the compact theory these appear as additional divergences in two and three point 
functions involving the mode of $\phi$ that is constant along the torus, $\phi_0$. 
The correct interpretation of these divergences is unusual dynamics for $\phi_0$. 
Since the mass of $\phi_0$ diverges as the ultraviolet cutoff is taken to infinity, 
$\phi_0$ no longer appears as a propagating degree of freedom in the low energy 
effective theory. 

In section 3 we discuss the Kaluza-Klein mass formula for the scalar field theory 
introduced in section 2.  There are one loop corrections which are singular as $\theta$ 
goes to zero. For small $\theta$, they lead to a four dimensional spectrum of states
which differs qualitatively from the Kaluza-Klein spectrum of a commutative theory. 
The one loop correction to the dispersion relation for the Kaluza-Klein excitations 
resembles that of winding states in string theory \texa.

In section 4 we briefly discuss similar issues in a $U(1)$ gauge theory compactified 
on $R^{1,3}\times T_{\theta}^2$. Again we find that modes of the gauge field 
which are constant along the torus disappear from the low energy theory.
The six dimensional gauge field contains Kaluza-Klein excitations that are 
either four-dimensional vectors or four-dimensional scalars. One loop
corrections modify the spectrum of the four-dimensional scalars, however, the
spectrum of the four-dimensional vectors is unchanged.

Concluding remarks are given in section 5.

\newsec{Perturbation Theory for $\phi^3$ on $R^{1,3} \times T^2_\theta$}

The noncommutative scalar $\phi^3$ theory in six noncompact dimensions
is defined  
by the following action:
\eqn\lag{
S = \int d^6x \left( {1\over 2} (\partial \phi)^2 - {1\over 2} m^2
\phi^2 -{ \lambda \over 3!}  
\phi \star \phi \star \phi \right).}
The coordinates of commutative four dimensional Minkowski space are 
$x^0,x^1,x^2,x^3$. 
The coordinates $x^4$ and $x^5$ are noncommuting : 
\eqn\com{ [x^4,x^5] = i \theta .} 
Since the theory defined by \lag\ contains an infinite number of
higher derivative  
operators, naively one would expect it to be
nonrenormalizable. However, it is a  
renormalizable theory because all ultraviolet divergences can be cancelled by 
counterterms of the form appearing in the action \lag . Unlike
conventional field theories, the counterterms are not local operators
but the nonlocality is of the same type present in the tree level action.
The S-matrix elements of this theory are plagued with infrared divergences. 
For instance, the one loop correction to the two point function leads to a
term in the effective action of the form
\eqn\opi{
-\int d^6x {\lambda^2 \over 2 (4 \pi)^3} \phi {1\over \partial \circ
\partial}\phi,} 
where we have introduced the inner product $p \circ p =- p_M
(\theta^2)^{M N} p_N = \theta^2(p_4^2+p_5^2)$.  

The action in \lag\ must be augmented by cutoff dependent counterterms. In this 
paper we work at lowest nontrivial order in perturbation theory (i.e. order $\lambda^2$ 
in 1PI diagrams) and ignore the obvious classical instability in \lag . The one loop 
counterterms in the noncompact theory are
\eqn\Sct{
S_{ct} = -\int d^6x \bigg( {\lambda^2 \over 8 (4 \pi)^3}
(\Lambda^2 - m^2 {\rm ln}\, \Lambda^2 )\phi^2
+ {\lambda^2 \over 48 (4 \pi)^3}{\rm ln}\, \Lambda^2 (\partial \phi)^2
+ {\lambda^3 \over 48 (4 \pi)^3}{\rm ln}\, \Lambda^2 \phi \star \phi \star \phi \bigg) .}
The explicit form of the ultraviolet regulator $\Lambda$ will be discussed later.

In this paper, we will be concerned with the theory on a space with two 
noncommuting compact directions. The divergence structure of the 
noncommutative theory on a compact space is different from that of 
the theory on a noncompact space. The coordinates $x^4$ and $x^5$ are 
compactified on a rectangular torus with $0 \leq x^4,x^5 \leq  2 \pi R$. Momenta 
along the compact directions are quantized, $\vec{p} = \vec{n}/R$, where 
$\vec{n}=(n_4,n_5)$ are integers. For the mode with $\vec{n} \neq  0$, the
term in the one particle irreducible action shown in \opi\ is finite. It 
gives rise to an interesting modification of the Kaluza-Klein spectrum
that will be discussed in section 3.  For the mode with $\vec{n}=0$, \opi\ is infinite.

Given the special role played by $\vec{n} =0$ mode, it is very useful
to make the following definitions
\eqn\zm{\eqalign{
\phi_0 &= {1 \over (2 \pi R)^2} \int dx^4 dx^5 \phi  \cr
\bar{\phi} &= \phi - \phi_0 .}}
The $\phi_0$ field 
contains the mode with $\vec{n}=0$, $\bar{\phi}$ contains all modes with 
nonvanishing $\vec{n}$. In terms of these fields the action \lag\ is: 
\eqn\lagc{
S = \int d^6x \left( {1\over 2} (\partial \bar{\phi})^2 - {1\over 2}
m^2 \bar{\phi}^2 + 
{1\over 2} (\partial \phi_0)^2 - {1\over 2} m^2 \phi_0^2 -
{\lambda \over 3!}  \bar{\phi} \star \bar{\phi} \star \bar{\phi} - 
{\lambda \over 2}  \bar{\phi}^2 \phi_0
- { \lambda \over 3!}  \phi_0^3
\right).}
The products involving $\phi_0$ are ordinary products. This is because after integration 
the $\star$-product of three fields reduces to an ordinary product whenever one of the 
fields has vanishing momentum along noncommuting dimensions. No operator of the form  
$\int d^6 x \bar{\phi} \phi_0^2$ appears because it is forbidden by momentum conservation.  

\ifig\frules{Feynman rules for noncommutative $\phi^3$ on $R^{1,3} \times T^2_\theta$.
Solid lines are $\bar{\phi}$ quanta, dotted lines are $\phi_0$ quanta.
Momenta along noncompact directions are denoted by $p$, while $\vec{n}/R,\vec{k}/R$ 
and $\vec{m}/R$ are momenta along compact directions. The wedge product is defined to 
be $\vec{n} \wedge \vec{k} \equiv n_4 k_5 - n_5 k_4$.  }
{\epsfxsize6in\epsfbox{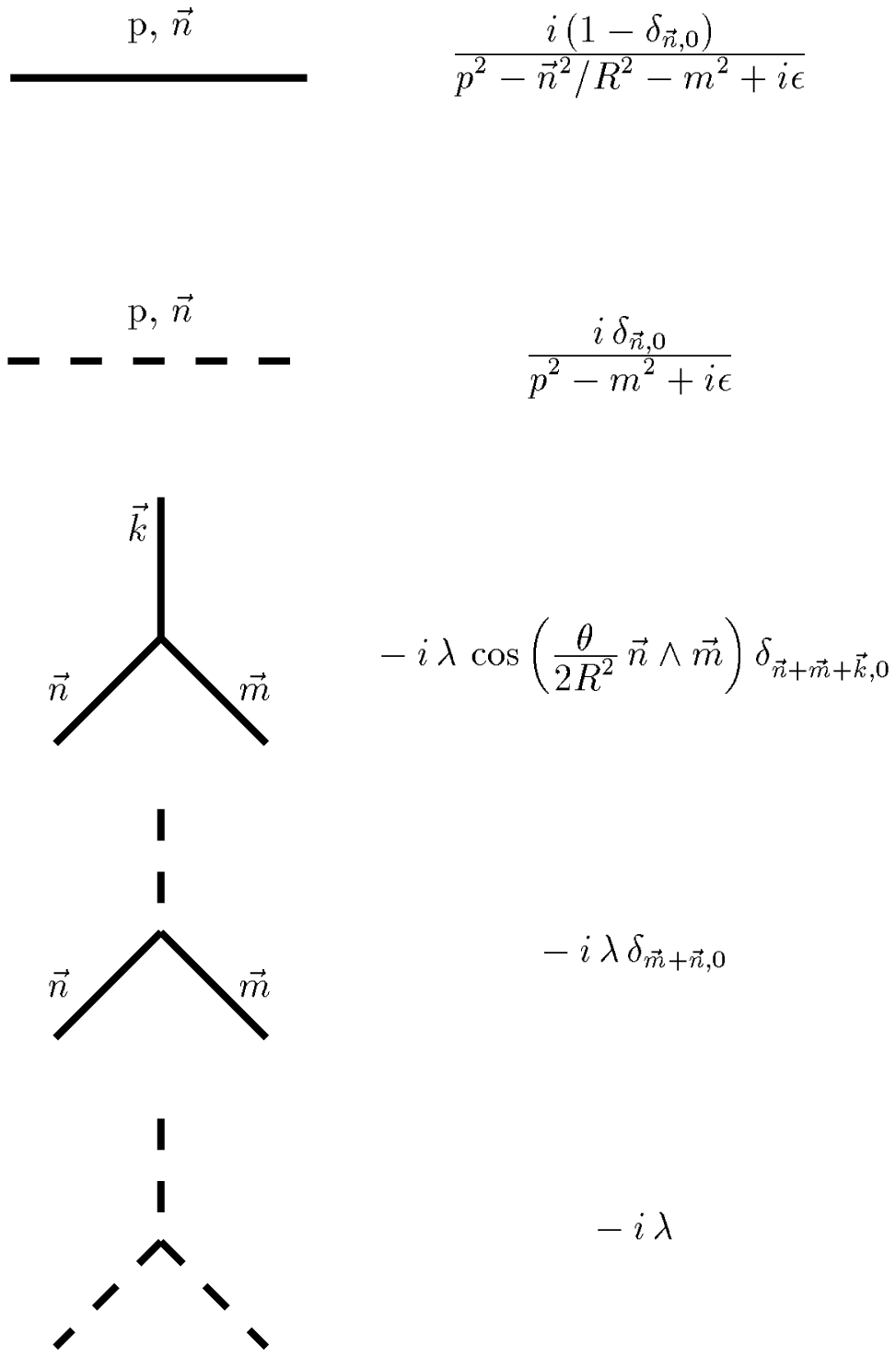}}

The Feynman rules derived from the action \lagc\ are given in \frules . Noncommutativity
introduces oscillatory factors in the vertices of the theory. These oscillatory factors never make a
graph more ultraviolet divergent than the naive power counting estimate.  Therefore if a graph 
is ultraviolet finite in the commutative theory it will also be ultraviolet finite in the 
noncommutative theory. Since $\phi^3$ in six dimensions is renormalizable, only the two and 
three point functions of this theory can have ultraviolet divergences.  

\ifig\twopt{One loop corrections to the two point functions for $\phi_0$ and $\bar{\phi}$.}
{\epsfxsize4in\epsfbox{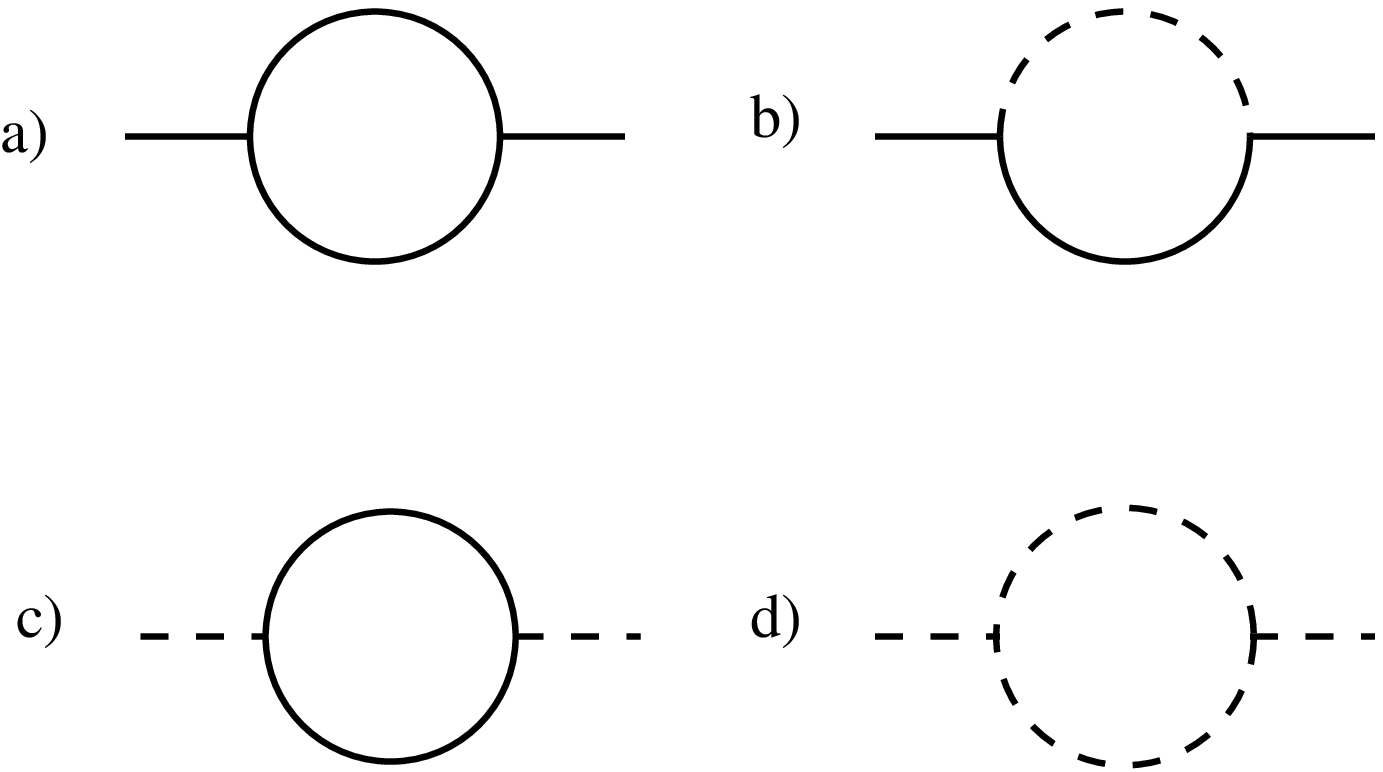}}

The one loop contributions to the two point functions are shown in \twopt . We begin
with diagram $a)$, which is a contribution to the two point function of $\bar{\phi}$. 
\eqn\tpa{
a) = {\lambda^2 \over 2} {1\over (2 \pi R)^2} \sum_{\vec{k}} \int {d^4 l \over (2 \pi)^4}
{\cos^2(\theta \,\vec{n} \wedge \vec{k}/(2 R^2) )(1-\delta_{\vec{k},0})(1-\delta_{\vec{n}+\vec{k},0})
\over (l^2 -\vec{k}^2/R^2 - m^2) ((l+p)^2 -(\vec{n}+\vec{k})^2/R^2 - m^2)} .}
In \tpa , $l$ denotes the loop momenta along the noncompact directions, while $\vec{k}/R$ 
is loop momenta along compact directions. Similarly, $p\, (\vec{n}/R)$ is the external 
momenta along the noncompact (compact) directions. In \tpa , $\vec{n} \wedge \vec{k} \equiv n_4 k_5 - n_5 k_4$. Since the external field is $\bar{\phi}$, $\vec{n} \neq 0$. 
This implies $\delta_{\vec{k},0}\delta_{\vec{n}+\vec{k},0} = 0$. We can simplify 
the numerator by using the half angle formula for the cosine and the fact that
$\vec{n} \wedge \vec{k} = 0$ for $\vec{k} = 0,-\vec{n}$ to obtain:
\eqn\tpat{
a) = {\lambda^2 \over 4} {1\over (2 \pi R)^2} \sum_{\vec{k}} \int {d^4 l \over (2 \pi)^4}
{1-2\, \delta_{\vec{k},0}-2\, \delta_{\vec{n}+\vec{k},0}+\cos (\theta\, \vec{n} \wedge \vec{k}/R^2 )
\over (l^2 -\vec{k}^2/R^2 - m^2) ((l+p)^2 -(\vec{n}+\vec{k})^2/R^2 - m^2) .}}
The oscillatory factor 
$\cos (\theta \, \vec{n} \wedge \vec{k}/R^2)$ makes the last term of \tpat\ ultraviolet finite.
The first term is quadratically divergent, while the second and third terms are logarithmically 
divergent. Only the divergent terms will be evaluated in this section.

We first represent the propagator using Schwinger parameters then perform the Gaussian 
integral over the loop momentum to obtain 
\eqn\aquad{\eqalign{
a) = {i \lambda^2\over 4(4 \pi)^3 \pi R^2} 
\int_0^\infty {d\alpha \over \alpha} \int_0^1 dx \bigg( \sum_{\vec{k}} \exp\left[
-\alpha \left(m^2 - x(1-x)p^2 +{\vec{k}^2 \over R^2} + x{\vec{n}^2 +2 \vec{n}\cdot\vec{k}
 \over R^2}\right)\right] \cr
- 4 \exp\left[-\alpha \left(m^2 - x(1-x)p^2 +x{\vec{n}^2 \over R^2} \right)\right] \bigg) .}}
The sum over $\vec{k}$ is performed using the definition of the Jacobi theta function
\eqn\jacobi{
\vartheta(\nu, \tau) = \sum_{n=-\infty}^{\infty} \exp(\pi i n^2 \tau +2 i\pi n \nu) ,} 
and the modular transformation $\vartheta(\nu,\tau)=(-i \tau)^{-1/2}\exp( -\pi i \nu^2/\tau) 
\vartheta(\nu/\tau, -1/\tau)$. The result is
\eqn\ansa{\eqalign{
{i \lambda^2\over 4(4 \pi)^3} \bigg\{ \int_0^\infty {d\alpha \over \alpha^2} \int_0^1 dx
 \exp\left[-\alpha \left( m^2 + x(1-x)\left(-p^2 +{\vec{n}^2 \over R^2}\right) \right) \right]
\hat{\vartheta}(x n_4) \hat{\vartheta}(x n_5) \cr
- {4 \over \pi R^2} \int_0^\infty {d\alpha \over \alpha} \int_0^1 dx
\exp\left[-\alpha \left( m^2 -x(1-x)p^2 +x {\vec{n}^2 \over R^2}\right) \right] \bigg\} .}}
In this formula we have defined $\hat{\vartheta}(\nu) = \vartheta(\nu, i \pi R^2/\alpha)$.
The ultraviolet divergent contribution comes from the $\alpha \rightarrow 0$ region and
in this limit we can set the $\hat{\vartheta}$ functions to unity. Then the first term is proportional to the 
quadratically divergent one loop correction in the commutative theory in 
the $R \rightarrow \infty$ limit. Corrections to this limit come 
from the $n \neq  0$ terms in the expansion of \jacobi\ and these are finite.  There 
are also corrections arising because of the absence of the $\vec{n}=0$ mode in the 
loop integral. These corrections depend explicitly on the size of the compact dimension and 
are logarithmically divergent since the loop integral is effectively 
four dimensional for the zero mode.

The integrals can be regulated  by inserting into the integral $\exp(-1/(\Lambda^2 
\alpha))$, where $\Lambda$ is the ultraviolet cutoff.  The divergent contribution from 
diagram $a)$ in \twopt\ is:
\eqn\resa{
a) = {i \lambda^2 \over 4(4 \pi)^3}\bigg[ \Lambda^2 - \bigg(m^2 
+ {1\over 6}\left(- p^2 + {\vec{n}^2 \over R^2}\right) + {4\over \pi R^2} \bigg){\rm ln} 
\, \Lambda^2 \bigg] .}
The divergences in the other graphs in \twopt\ can be computed in the same way. We find
\eqn\uvd{\eqalign{
b) =& {i \lambda^2 \over (4 \pi)^3}{{\rm ln} \, \Lambda^2 \over \pi R^2} , \cr
c) =& {i \lambda^2 \over 2(4 \pi)^3}\bigg[ \Lambda^2 - \bigg(m^2 
-{p^2\over 6} + {1\over \pi R^2} \bigg){\rm ln} \, \Lambda^2 \bigg] , \cr
d) =& {i \lambda^2 \over 2(4 \pi)^3}{{\rm ln}\, \Lambda^2 \over \pi R^2} .}}

\ifig\threept{Divergent one loop corrections to the three point functions.}
{\epsfxsize4in\epsfbox{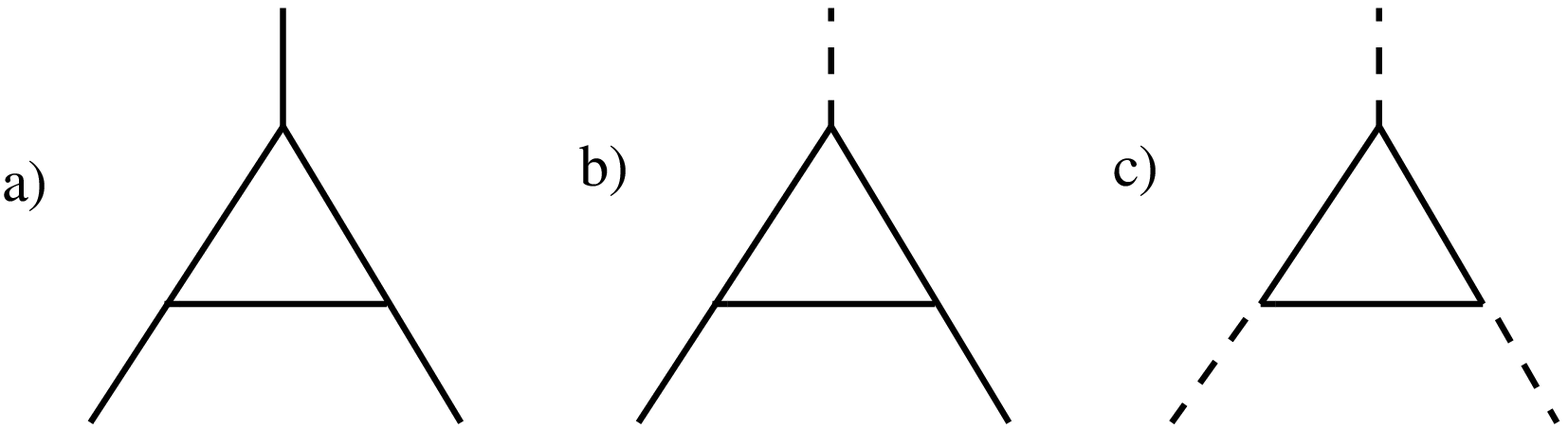}}

Next we consider the one loop divergences in the three point functions of the theory. 
If a zero mode appears in the loop then the loop integration is effectively four 
dimensional and the graph is finite by power counting. Therefore, it is only necessary 
to consider loop graphs with internal $\bar{\phi}$ quanta. The relevant Feynman graphs 
are shown in \threept .  The leading divergences are logarithmic and these can be
obtained from a calculation treating the loop momentum as continuous. The divergences 
in the graphs of 
\threept\ are
\eqn\threuv{\eqalign{
a) =&  - {i \lambda^3 \over 8 (4 \pi)^3} 
\cos \left( {\theta \over  2 R^2} \vec{n} \wedge \vec{k}\right) {\rm ln}\, \Lambda^2, \cr
b) =&  - {i \lambda^3 \over 4 (4 \pi)^3}  {\rm ln}\, \Lambda^2, \cr
c) =& - {i \lambda^3 \over 2 (4 \pi)^3} {\rm ln}\, \Lambda^2 .}}

The two and three point functions for $\bar{\phi}$ are rendered finite by adding the 
counterterms in \Sct .  These counterterms do not render correlation
functions involving $\phi_0$ finite in the compactified theory. After adding the counterterms
in \Sct, the divergent part of the $\phi_0$ self energy is
\eqn\signot{
\Sigma_0 = - {\lambda^2 \over 4(4 \pi)^3 } \left[ \Lambda^2 - \left( m^2 -{p^2 \over 6}
\right){\rm ln}\, \Lambda^2 \right] .}
A similar uncancelled divergence appears in three point functions involving $\phi_0$.
The counterterm in \Sct\ regulates three point involving three $\bar{\phi}$ but not the
divergences from diagrams $b)$ and $c)$ in \threept .

\ifig\resum{Resummation of divergent terms in $\phi_0$ two-point function. The shaded
blob represents $\Sigma_0$ in \signot .}
{\epsfxsize4in\epsfbox{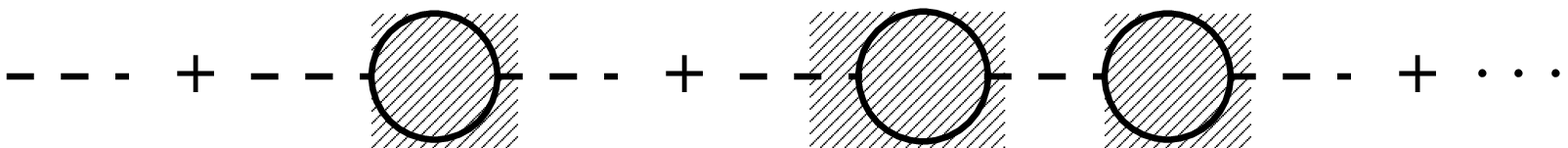}}

In the noncompact theory the two and three point functions contain infrared divergences. 
In the compactified theory, these infrared divergences appear
as uncancelled cutoff dependence in two and three point functions involving $\phi_0$.
The correct interpretation of the uncancelled dependence on $\Lambda$ is obtained by 
considering renormalization from the Wilsonian point of view.  Instead of trying to introduce 
new counterterms, we interpret the ultraviolet cutoff as a physically significant scale. 
Then it is clear that the $\phi_0$ mode, unlike the $\bar{\phi}$ modes, develops a mass of 
order $\Lambda$ and hence decouples from the low energy theory as  $\Lambda \rightarrow 
\infty$.  This decoupling can be seen diagramatically. As $\Lambda \rightarrow \infty$, 
$\Sigma_0 \rightarrow \infty$. When $\Sigma_0$ becomes large it is necessary sum insertions 
of $\Sigma_0$ in the $\phi_0$ propagator to all orders, as shown in \resum .  The two-point 
correlation function $\langle 0 | T(\phi_0(x) \phi_0(y))| 0 \rangle$ is of order $1/\Lambda^2$  
and hence vanishes as $\Lambda^2 \rightarrow \infty$. Other correlation functions involving
$\phi_0$ also vanish.

In any correlation function, the resummation shown in \resum\ is to be performed whenever
$\phi_0$ is an external line or for graphs which fall apart when a $\phi_0$ propagator is cut. 
Furthermore, at the order in perturbation theory we are working this resummation is only 
performed for $\phi_0$ propagators which do not appear inside 1PI subgraphs. For example, 
this resummation is not performed for internal $\phi_0$ in  diagrams $b)$ and $d)$ of \twopt . 
This procedure gives finite answers for all correlation functions at this order in
perturbation theory. There are 
uncancelled divergences proportional to ${\rm ln}\,\Lambda^2$ in the 1PI three point functions 
with one or three external $\phi_0$. However, the resummed propagators for the $\phi_0$ fall off as
$1/\Lambda^2$. 

Because of the sign of the correction in \signot , the mass of the zero mode
appears to be driven to negative, perhaps indicating that $\phi_0$ is tachyonic. Unfortunately,
it is impossible to address this issue in the one loop approximation we are employing in
this paper. This is because higher order corrections to the self energy are of the form 
\eqn\hoc{
\delta \Sigma_0 \sim  \lambda^{2n +2} \Lambda^2 {\rm ln}^n\Lambda .}
Perturbation theory breaks down for $\Lambda \rightarrow \infty$ and the sign of
the quadratic divergence cannot be determined by lowest order perturbation theory.
The zero mode mass is likely to be driven to $\pm \infty$,
but the sign is not known from the one loop calculation we have performed in this section. 
At the next order in perturbation theory $\Sigma_0$ will occur in loops that renormalize the
masses of the $\bar{\phi}$ modes. The cutoff dependence of these corrections must be cancelled 
by the counterterms if the low energy effective theory has dynamics for the modes of $\bar{\phi}$ 
that is close to what is expected based on the one loop analysis.

This completes the analysis of the cutoff dependence of the compactified theory at the one 
loop level.  All necessary counterterms are present in \Sct\ and the additional divergences 
remove $\phi_0$ from the low energy theory. An important outstanding issue is to see how 
ultraviolet divergences cancel at higher orders.

\newsec{Kaluza-Klein Spectra on Noncommutative Tori}

In this section we compute the  spectrum of Kaluza-Klein modes that arises from
\lag\ compactified on $T^2_\theta$. In the classical limit, the $\phi$ quanta  
of the commutative and noncommutative theories have the same dispersion relation. 
However, the nonlocality of noncommutative field theories leads to one loop
corrections to the dispersion relation. In the noncompact case the one loop 
dispersion relation is of the form
\eqn\mdr{ 
p^2 -m^2 +{\lambda^2  \over (4\pi)^3}{1 \over p \circ p} + . . . = 0,}
where the ellipsis denote terms that are less singular as $p \circ p \rightarrow 0$.
The nonanalytic dependence in $\theta$ of \mdr\ results in corrections that diverge for 
soft momenta along the noncommutative directions.  After compactification, the 
discretization of momenta isolates the zero momentum mode of the field, such that only 
for this component of the field one obtains a divergence in the one loop dispersion 
relation \mdr . As shown in section $2$ this quantum correction drives the $\phi_0$ mass
to infinity, effectively removing $\phi_0$ from the theory. 

It is straightforward to compute the one loop dispersion relation for the non-zero modes. The 
relevant Feynman graphs that need to be evaluated are diagrams $a)$ and $b)$ of \twopt . As 
shown in section $2$, the divergences in these graphs are absorbed by counterterms. 
We are interested in the finite corrections coming from the nonplanar piece of diagram $a)$,
i.e., the term proportional to $\cos(\theta\,\vec{n}\wedge\vec{k}/R^2)$ in \tpat .
This piece gives the leading behavior for small $\theta$. The one loop self energy is 
\eqn\nonpla{\eqalign{
\Sigma=-{\lambda^2\over 4(4\pi)^3}\int_0^1\ dx\int_0^\infty\
d\alpha 
\ \alpha^{-2}\exp\left[-\alpha\left(m^2+x(1-x)
\left(-p^2 + {\vec{n}^2 \over R^2}\right) \right)-
{\theta^2\vec{n}^2\over 4R^2\alpha}\right] \cr
\times\bigg\{{1\over 2}\hat{\vartheta}\left(xn_4+i{\theta n_5 \over 2\alpha}\right)
\hat{\vartheta}\left(xn_5-i{\theta n_4 \over 2\alpha} \right)  
+ {1\over 2}\hat{\vartheta}\left(xn_4-i{\theta n_5 \over 2\alpha}\right)\hat{\vartheta}
\left(xn_5+i{\theta n_4 \over 2\alpha} \right) \bigg\} .}} 
Setting $p^2-\vec{n}^2/R^2 = m^2$ in the above formula, 
the leading terms for small $\theta$ are
\eqn\expres{
\Sigma =-{\lambda^2\over (4\pi)^3}\left({R^2\over \theta^2 \vec{n}^2}+{5 \over 24}
m^2 \ln\left({m^2\theta^2\vec{n}^2 \over R^2}\right)+ ...\right),}
so that the formula for the Kaluza-Klein masses is
\eqn\fol{
m_{\vec{n}}^2 = m^2 + {\vec{n}^2 \over R^2} -{\lambda^2 \over (4\pi)^3}
\left( {R^2 \over
\vec{n}^2 \theta^2} +{5 \over 24} m^2\ln\left({m^2\theta^2
\vec{n}^2 \over R^2}\right)+ ...\right),}
Spatial noncommutativity in the compact directions gives a correction to the 
Kaluza-Klein mass formula which resembles that of winding states in string theory.
Note that the mass correction from the nonplanar diagram is
negative. For sufficiently small $\theta$, this correction 
induces a negative mass squared for some Kaluza-Klein modes. The ellipses
in Eqs. \expres\ and \fol\ denote terms that are less important as $\theta 
\rightarrow 0$. 

\newsec{Noncommutative $U(1)$ Gauge Theory in Six Dimensions}

In a noncommutative space-time a pure $U(1)$ gauge theory is
interacting. The action in six dimensions is 
\eqn\action{
S = - {1 \over 4} \int d^6 x F_{MN}\star F^{MN} ,}
where the field strength tensor 
\eqn\fs{
F_{MN}  = \partial_M A_N - \partial_N A_M - i g (A_M \star A_N -A_N
\star A_M),} 
transforms as 
\eqn\transform{
\delta_\alpha F_{M N} =  i g (\alpha \star F_{M N} - F_{M N} \star \alpha) }
under the $U(1)$ gauge transformation
\eqn\gt{
\delta_\alpha A_M = \partial_M \alpha + i g(\alpha \star A_{M} - A_{M}
\star \alpha).} 
Terms in the action quadratic in $A$ are the same as in the
commutative theory so the  
tree level spectrum of Kaluza-Klein excitations follows that case.  
These are massive four vectors, labelled by
$A_{\mu}^{(\vec{n})}$,  and scalars, labeled by 
$h^{(\vec{n})} = (n_5 A_4^{(\vec{n})} - n_4 A_5^{(\vec{n})})/\sqrt{\vec{n}^2}$,
with masses 
\eqn\mass{
m_{\vec{n}}^2 = {\vec{n}^2 \over R^2} .}
Note that the orthogonal scalar component,
$g^{(\vec{n})} = (n_4 A_4^{(\vec{n})} + n_5 A_5^{(\vec{n})})/\sqrt{\vec{n}^2}$,
can be gauged away. It has become the longitudinal component of the
massive four 
vectors $A_{\mu}^{(\vec{n})}$ through the standard Higgs mechanism.

As in the case of scalars discussed in Section 2 the $\vec{n}=0$ modes need to be 
treated differently. They consist of a four dimensional gauge field and two massless 
scalars. In this case the zero mode fields do not interact with each other nor do they 
couple to the $\vec{n} \neq 0$ modes. There are no one loop graphs contributing to the two point
functions of these fields in the compactified theory. However, these fields receive a divergent 
wavefunction renormalization from the counterterms of the noncompact theory. This divergence 
can be removed simply by making a field redefinition. 
If currents charged under the U(1) gauge symmetry were added to the theory, the field 
redefinition would result in a suppression of the couplings of the gauge field to these currents.
The $\vec{n}=0$ modes decouple from the low energy effective field theory,
not because their masses are driven to infinity (as in the case of the scalar theory), 
but because their couplings to other fields are driven to zero.

The perturbative corrections to the masses of the Kaluza-Klein modes with 
$\vec{n} \neq 0$ follow from the two point 
function $\Pi_{MN}$ of the gauge field. As in the case of the 
scalar field theory considered 
in the last section, the most important corrections for $\theta < R^2$
can be obtained by replacing the discrete sum over compact loop
momenta with a continuum integral. This yields
\nref\gkmrs{J.Gomis, M. Kleban, T. Mehen, 
M. Rangamani and S. Shenker, ``Noncommutative Gauge Dynamics from the 
String World Sheet'', hep-th/0003215.}%
\sunew\haya\gkmrs
\eqn\selfenergy{
\Pi_{M N} =\Pi^p_{M N} + \Pi^{np}_{M N} ,}
where \foot{$\Pi_{M N}$ is calculated in background field gauge. This amounts
to replacing $A_M$ in Eq.\ action\ by $Q_M+A_M$ and adding the gauge fixing term
$1/2 \int d^6x G^2$, where $G = \partial_M Q^M - i g(A_M \star Q^M -Q_M \star A^M)$.
Note that we are using the convention $g_{\mu \nu} = {\rm diag}(+,-,-,-,-,-)$.}
\eqn\pln{
\Pi^p_{M N} =  {i g^2 \over (4 \pi)^3}\int_0^{\infty}{d \alpha \over \alpha^2} \int_0^1
dx  \exp[p^2 \alpha  x(1-x)] \left[ (p^2 g_{M N} - p_M
p_N)(8-4(1-2 x)^2)  \right]  ,}
and 
\eqn\nplan{\eqalign{
\Pi^{np}_{MN} =  -{i g^2 \over (4 \pi)^3}\int_0^{\infty}{d \alpha \over \alpha^2} 
\int_0^1 dx &
\exp\left(p^2 \alpha  x(1-x) -{p \circ p \over 4 \alpha}\right) \cr
& \times \left[ (p^2 g_{MN} - p_M
p_N)(8-4(1-2 x)^2)  
- {4 \over \alpha^2}\tilde{p}_M \tilde{p}_N \right]  .}}
Here $\tilde{p}_M = \theta_{M N} p^N$ and $p$ is the six dimensional external 
momentum. The components of external momentum along the compact directions, 
$p_4$ and $p_5$, are quantized.  

The first term inside the brackets in \nplan\ gives rise to
wavefunction renormalization 
for the massive vector modes $A^{(\vec{n})}_{\mu}$ and does not change
their masses. 
However, the term proportional to $\tilde{p}_M  \tilde{p}_N$ can
affect the masses of the  
scalars $h^{(\vec{n})}$. Using the same approximations that were used
in the scalar case we find 
that the  Kaluza-Klein spectrum for the $h^{(\vec{n})}$ at one loop is
\eqn\hkk{
m_{\vec{n}}^2 = {\vec{n}^2 \over R^2} - {8 g^2 R^4 \over \pi^3
\theta^4 (\vec{n}^2)^2} + ... \, .} 
Again for small $\theta$ the second term in \hkk\ can dominate over 
the first and drives $m_{\vec{n}}^2$ negative. Higher order terms in the
effective potential for $h^{(\vec{n})}$ must be computed to determine
whether this results in a vacuum expectation value for the $h^{(\vec{n})}$ or signals a
true instability of the theory. 

\newsec{Conclusion}

It is a widely held belief that there are additional compactified spatial
dimensions. Extra dimensions occur in string theory and it has even been
conjectured that their presence may lead to a resolution of the hierarchy
puzzle \hierarchy\ or cosmological constant \cc\ problem. If extra spatial dimensions do exist
then it is possible that they are noncommuting. Motivated by this possibility
we have explored the properties of some simple theories with noncommuting
compact extra dimensions.

We considered, at lowest order perturbation theory, a $\phi^3$ theory compactified on a
rectangular noncommuting two torus, $T^2_{\theta}$. The UV-IR mixing that occurs in the
uncompactified case has important implications for the compactified theory. It causes the 
mode of the $\phi$ field that is constant on the torus (i.e. the zero mode) to have different 
dynamics than the other modes. Quantum corrections drive correlation 
functions of this mode to zero. It does not appear as a propagating degree of freedom in the 
effective four dimensional theory. 

The mass spectrum of Kaluza-Klein excitations was calculated in this theory.
At the one loop level the masses get a contribution that resembles that of 
winding states in string theory. This contribution  to the square of the mass 
is negative and for small values of the noncommuting parameter it could 
dominate over the tree level term.

We also examined a U(1) gauge theory compactified on $T^2_{\theta}$. Similar
phenomena to those that occur in the scalar case were found.

\centerline{\bf Acknowledgments}

We would like to thank Hirosi Ooguri for reading an earlier draft of this paper and 
Edward Witten for useful comments. J.G. would like to thank CERN for hospitality 
during the final stages of this work.  J.G. and T.M. are supported in part by the DOE
under grant no. DE-FG03-92-ER 40701.

\listrefs

\end